\documentclass{ethpaper}
\usepackage{graphicx}
\usepackage{here}
\usepackage{amssymb}
\usepackage{amsmath}
\usepackage{subfigure}
\usepackage{lineno}

\begin{document}
\begin{titlepage}
\ethnote{}
\title{Response evolution of the CMS ECAL\\
and R\&D studies\\
for electromagnetic calorimetry\\at the High-Luminosity LHC}
\begin{Authlist}
F.~Nessi-Tedaldi
\Instfoot{eth}{Institute for Particle Physics, ETH Zurich, 8093 Zurich, Switzerland}
on behalf of the CMS Collaboration
\end{Authlist}
\maketitle
\begin{abstract}
While the CMS experiment is currently harvesting LHC collision data at CERN, the performance of its electromagnetic calorimeter (ECAL) is being constantly monitored, and work has started to assess the need for changes to the detector to ensure an adequate performance for High-Luminosity LHC (HL-LHC) running, which is planned for 2022 and beyond.

In this paper, results from CMS running, beam tests and laboratory measurements are combined to anticipate the detector performance evolution at the HL-LHC. Further, various R\&D studies are illustrated, that will provide a useful choice for electromagnetic calorimetry at the HL-LHC.
\end{abstract}
\vspace{7cm}
\conference{\em To be published in the 2012 IEEE Nuclear Science Symposium Conference Record\\
also available as CMS Conference Report CR-2012/296 }
\end{titlepage}
\section{Introduction}
The detectors now in operation at the Large Hadron Collider (LHC) at CERN  will have to
face a challenging environment after the accelerator upgrade to
High-Luminosity running (HL-LHC), that is currently being planned to start in 2022.
The radiation levels expected there are typically a factor 10 larger, and the integrated luminosities ---  thus integrated radiation levels --- a factor 6 more important than the design values at the LHC. This scenario also applies to the CMS electromagnetic calorimeter (ECAL), which was designed for the initially  planned 10 years of LHC running in pp collisions at 14 TeV with a peak luminosity of $10^{34}\;\mathrm{cm^{-2}s^{-1}}$, up to a total integrated luminosity of $5\times 10^5\;\mathrm{pb^{-1}}$\cite{r-ECALTDR}.

In view of the HL-LHC running scenario, basic studies on Lead Tungstate  ($\mathrm{PbWO}_4$) crystals~\cite{r-LTNIM,r-pionNIM,r-LYNIM}  have established that high-energy protons and pions  cause a
cumulative loss of light transmission in PbWO$_4$, which is permanent at room temperature,
while  no hadron-specific damage to the scintillation emission is observed.
The cumulative nature of the damage  implies that the light output from Lead Tung\-state calorimeter crystals is expected to
decrease~\cite{r-RADQ,r-UTP,r-SCINT11,r-SING} in proportion to their integrated exposure to energetic hadrons, possibly beyond the level required for an optimum performance. Since a change of light output is expected to affect the achievable energy resolution, the tolerable amount of damage will depend on the physics goals of the experiment.

With the currently available amount of information, it is possible to attempt extrapolating the 
light output loss  as a function of integrated luminosity expected for the CMS electromagnetic calorimeter in its present configuration.
Further, CMS collision data, beam and laboratory test results have been combined to quantify the achievable energy resolution for electrons and photons as a function of hadron fluence, to anticipate a possibly unacceptable performance degradation.
In parallel,  R\&D work is being performed to identify suitable materials and optimize a calorimeter design for HL-LHC running conditions.
This paper addresses the status of all these studies, aimed at preparing an optimum upgrade of the CMS ECAL end caps, were it to be needed.

\section{The electromagnetic calorimeter of CMS}
\label{ECAL}
The CMS ECAL is a compact, hermetic, fine-grained and homogeneous calorimeter made of 75848
lead-tungstate scintillating crystals, arranged in a quasi-projective geometry and
distributed in a barrel region (EB), with pseudorapidity coverage up to $\lvert\eta\rvert = 1.48$, closed by
two endcaps (EE) that extend its coverage up to $\lvert\eta\rvert=3$. The scintillation light is readout with avalanche
photodiodes (APDs) in the EB and with vacuum phototriodes (VPTs) in the EE. To facilitate photon
identification, the crystals have a transverse size comparable to the Moli\`ere radius ($R_M= 21$ mm) in  $\mathrm{PbWO}_4$.
The front-face cross section of the crystals is $22\times 22\;\mathrm{ mm}^2$ in the EB and
$28.6\times 28.6\;\mathrm{mm}^2$  in the EE, with crystal depths of 26 and 25 radiation lengths, respectively.
Preshower detectors (ES) comprising two planes of silicon sensors interleaved with lead absorber
for a total of three radiation lengths are located in front of each endcap, at $1.65 < \lvert\eta\rvert <  2.6$,
to help in $\pi^0/\gamma$ separation. Electron/photon separation is limited to the region covered by the
silicon tracker ($\lvert\eta\rvert <  2.5$), which defines the acceptance region for photons in the $\mathrm{H}\rightarrow\gamma\gamma$ search.

The CMS ECAL is key to high-precision electron and photon calorimetry in CMS, and also constitutes the first nuclear interaction length of the CMS hadron calorimeter. It has already played a crucial role in the discovery of a Higgs-like particle
with the CMS detector~\cite{r-CMSHGG,r-NSS12KT,r-NSS12AT}.
\section{The environment}
\label{environment}
While the LHC Collider has already reached a centre-of-mass energy $\sqrt{s}=8$ TeV, an instantaneous peak luminosity ${\cal L}  = 7.7 \times 10^{33}\mathrm{cm}^{-2}\mathrm{s}^{-1}$ and an integrated luminosity of $\int{\cal L } dt = 20 \;\mathrm{fb}^{-1}$, during the long shutdown (LS1) planned for the period between mid february 2013 and the end of 2014,  an increase in energy up to 13 - 14 TeV is planned. It is then envisaged to increase further the peak luminosity through an injection upgrade in 2018 (LS2 shutdown), and the actual HL-LHC running would be attained after a third long shutdown (LS3) in 2022-2023, where the LHC interaction region upgrade would allow for a peak luminosity of 
${\cal L}  = 10^{35}\mathrm{cm}^{-2}\mathrm{s}^{-1}$ and a leveled value of ${\cal L}  = 5 \times 10^{34}\mathrm{cm}^{-2}\mathrm{s}^{-1}$, so that towards 2030, an integrated luminosity of 3000 $\mathrm{fb}^{-1}$ can be ultimately reached.
The nominal end of LHC running would be attained just before LS2, with 500 $\mathrm{fb}^{-1}$ of data collected, as originally planned~\cite{r-xxLHCTDR}.

While the LHC detectors were designed to withstand the corresponding radiation levels~\cite{r-CMSTP}, for HL-LHC running, after LS3, they would need to perform in an environment where the instantaneous radiation levels and particle fluxes are 10 $\times$ higher than in the LHC, and the integrated radiation levels and particle fluences would reach a factor 6 the LHC values.
The CMS detector, in particular, was designed~\cite{r-CMSTP} to perform adequately through 10 years of running, for ${\cal L}  = 1 \times 10^{34}\mathrm{cm}^{-2}\mathrm{s}^{-1}$ and  $\int{\cal L } dt = 500 \;\mathrm{fb}^{-1}$. For its electromagnetic calorimeter, this corresponds to ionizing radiation levels, at shower maximum, of 0.3 Gy/h in the EB, and 6.5 Gy/h in the EE at $\lvert\eta\rvert=2.6$, and to energetic hadron fluences ($> 20$ MeV) of $4\times 10^{11}\;\mathrm{cm}^{-2}$ in the EB and 
$3\times 10^{13}\;\mathrm{cm}^{-2}$ in the EE at $\lvert\eta\rvert=2.6$~\cite{r-ECALTDR}.
The corresponding levels during HL-LHC running with a peak luminosity ${\cal L}_{\mathrm{max}}  = 1 \times 10^{35}\mathrm{cm}^{-2}\mathrm{s}^{-1}$, would be, for ionizing radiation at shower maximum, 3 Gy/h in the EB, and 65 Gy/h in the EE at $\lvert\eta\rvert=2.6$, and after  $\int{\cal L } dt = 3000 \;\mathrm{fb}^{-1}$ of delivered integrated luminosity, energetic hadron fluences of $2\times 10^{12}\;\mathrm{cm}^{-2}$ in the EB and 
$2\times 10^{14}\;\mathrm{cm}^{-2}$ in the EE at $\lvert\eta\rvert=2.6$~\cite{r-CMSUTDR}.

It is evident, that to envisage HL-LHC running, the detector performance evolution has to be evaluated and, if needed, upgrades have to be foreseen.

\section{ECAL response variations during LHC running}
\begin{figure}[!t]
\centering
\includegraphics[width=3.5in]{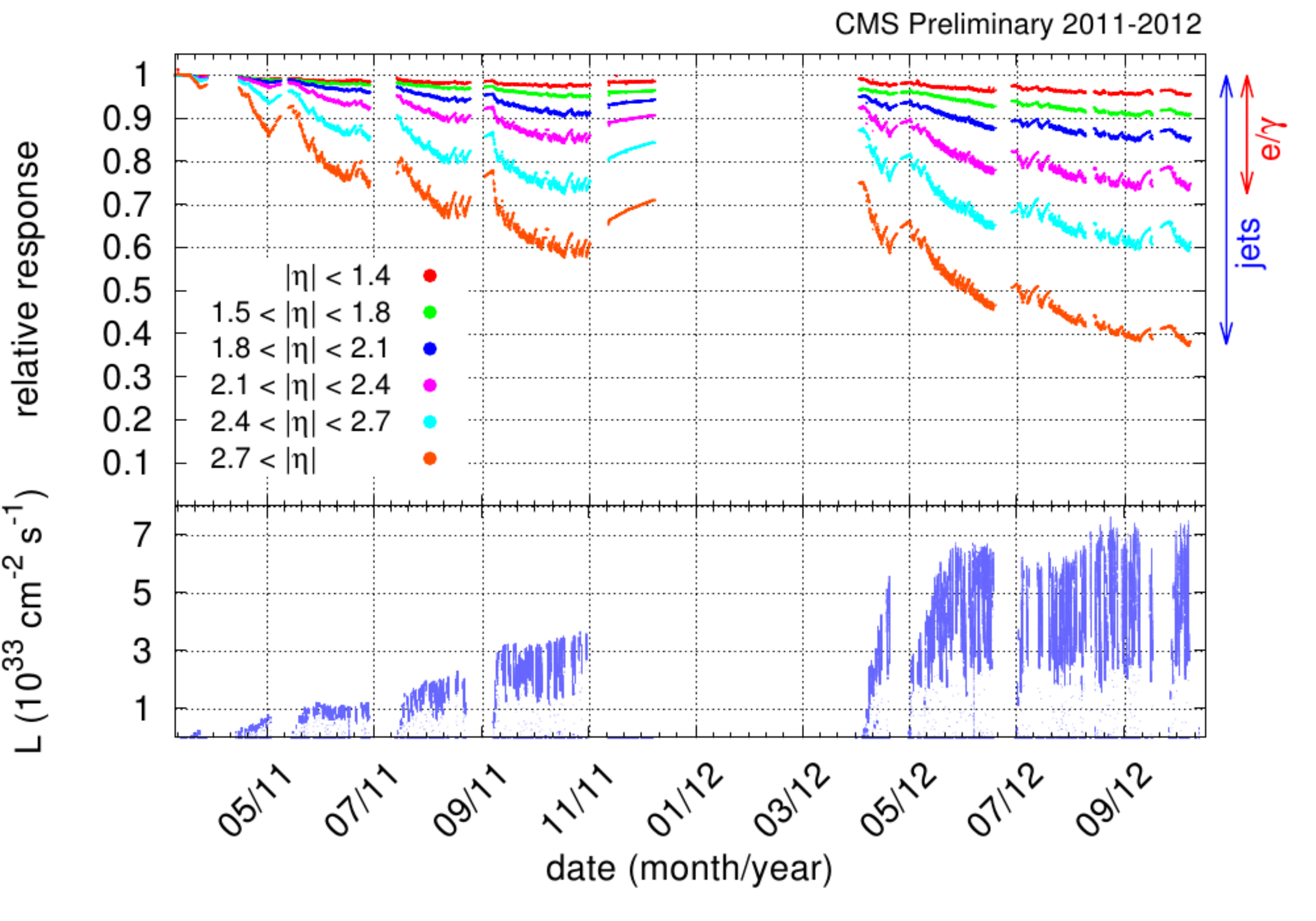}
\caption{Top: CMS ECAL relative response to laser light (440 nm) versus time, averaged over all crystals in one given pseudorapidity bin, for the
2011 and 2012 data taking. Bottom: Instantaneous luminosity versus time. }
\label{f-MON}
\end{figure}

The ECAL being a homogeneous crystal calorimeter, its response can be monitored through light injection~\cite{r-NSS12DL,r-NSS12KZ}. The changes observed during 2011-2012 running are visible in Fig.~\ref{f-MON}.
They are of the order of a few percent in the EB, in agreement with the values obtained during Quality Assurance at the time of construction, where losses $<6$\% for dose rates under $0.15$ Gy h$^{-1}$ were certified. The changes reach 25\% in the most forward EE regions used for electron and photon reconstruction. The response change is up to 60\% in channels closest to the beam pipe, were the exposure to ionizing radiation is highest.
\begin{figure}[!b]
\centering
\includegraphics[width=3.5in]{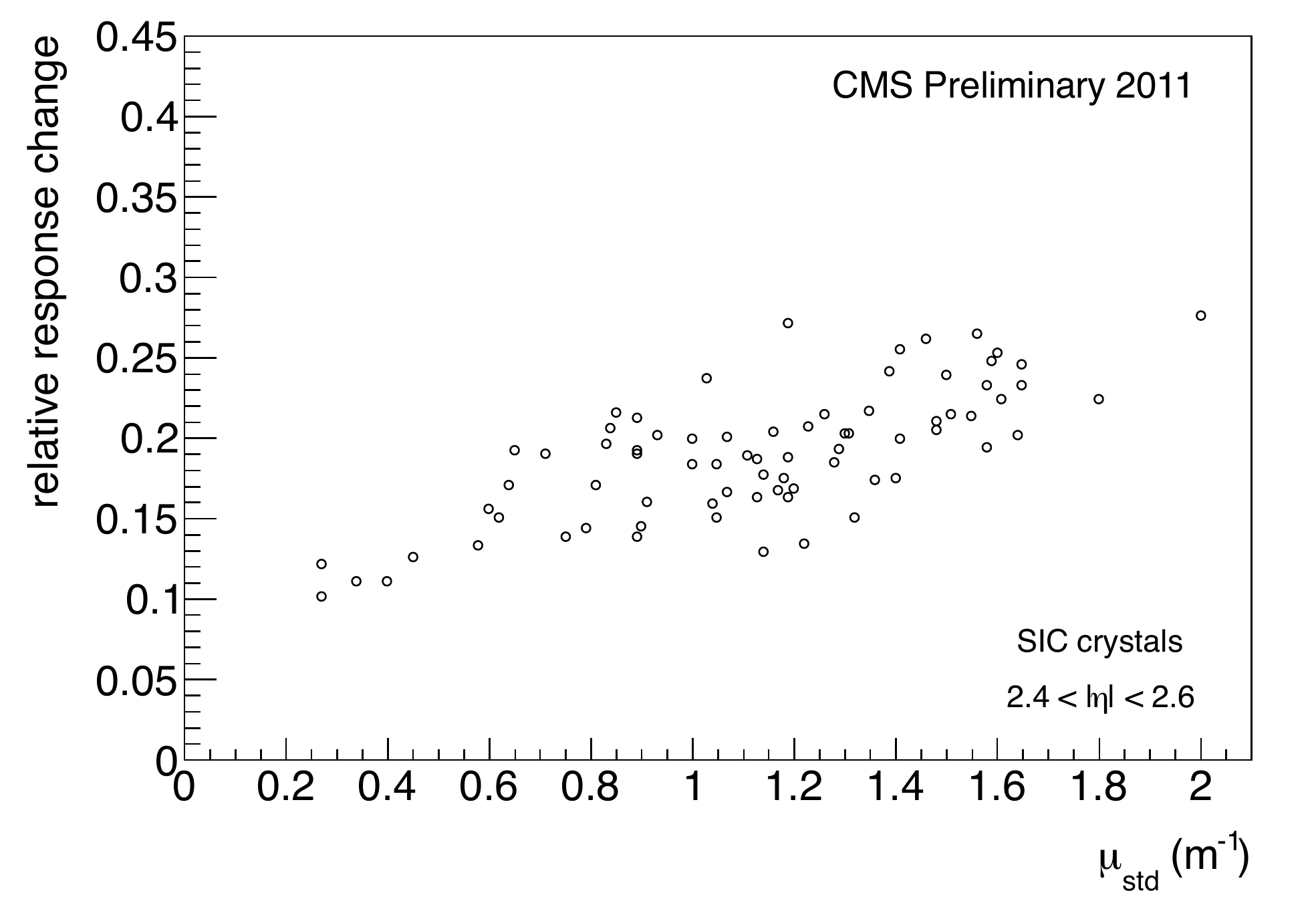}
	\caption{Correlation between relative change of response from ECAL monitoring data, observed between September 11th, 2011 and October 4th, 2011, and the absorption coefficient $\mu_{\mbox{std}}$ from a standard irradiation performed during Quality Control (one entry per crystal)~\cite{r-RADQ}.}
\label{f-HJW}
\end{figure}

The crystals undergo cycles of loss and recovery, over a few hours, that are mostly due to transmission losses from ionizing radiation damage. The reason lies in the well-known radiation-induced creation of color centers which absorb the scintillation light~\cite{r-RADQ} and which recover during periods without beam or with extremely low luminosity. All the effects observed in 2011 and 2012 running are consistent with radiation hardness tests performed during construction, in that no EB channels show losses beyond the expectations, while larger losses are observed in the EE, also possibly due to other sources of change.

Work is in progress to disentangle various possible sources of signal changes in the EE, one example of which can be appreciated in Fig.~\ref{f-HJW}.
There, the correlation is shown, for a number of crystals situated in the region $2.4<\lvert\eta\rvert<2.6$, between relative change of response from ECAL monitoring data, and the absorption coefficient $\mu_{\mbox{std}}$ from a standard irradiation performed during Quality Control~\cite{r-RADQ}. It shall be recalled here that the induced absorption coefficient $\mu_{IND}(\lambda)$as a function of light wavelength $\lambda$ is
defined as:
\begin{equation}
\mu_{IND}(\lambda) = \frac{1}{\ell}\times \ln \frac{LT_0 (\lambda)}{LT (\lambda)}
\label{muDEF}
\end{equation}
where $LT_0\; (LT)$ is the Longitudinal Transmission value measured
before (after) irradiation through the length $\ell$ of the crystal.
 The evident correlation proves that a fraction of the signal loss is due to a change in crystal transmission from ionizing damage, in that channels with radiation harder crystals (smaller $\mu_{\mbox{std}}$ values) exhibit a smaller loss amplitude than channels where crystals are less radiation hard, and thus were qualified with a higher $\mu_{\mbox{std}}$. The non-vanishing intercept, however, indicates the presence of a fraction of signal loss, uncorrelated with crystal radiation hardness, which are caused by other sources, like, e.g., losses in VPT response and possibly early cumulative losses from hadron-specific changes in crystal transmission~\cite{r-LTNIM}, as discussed in the following section.
\section{The CMS ECAL at the HL-LHC}
\label{s-HL-LHC}
\begin{figure}[!b]
\centering
\includegraphics[width=3.5in]{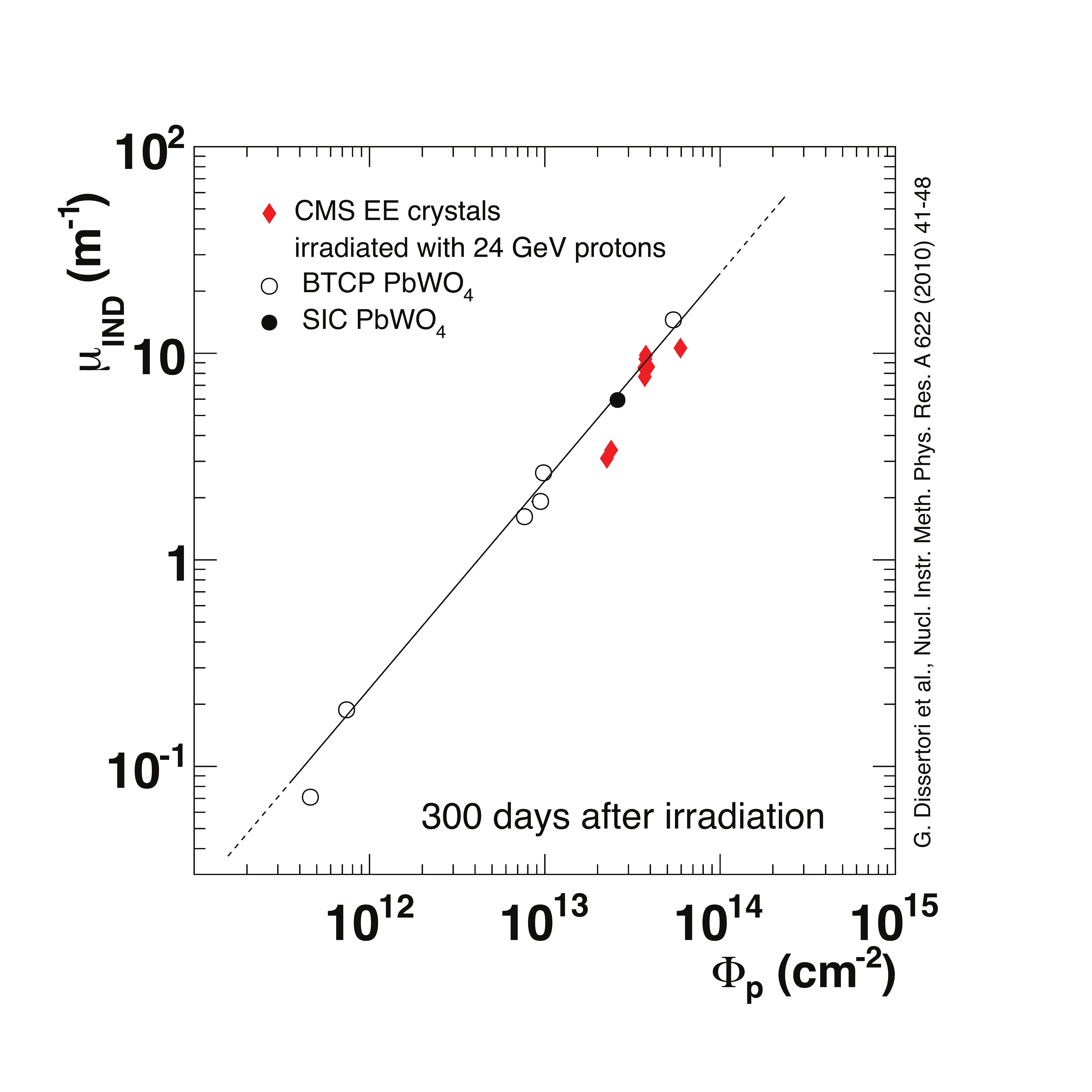}
	\caption{Induced absorption coefficients versus proton fluence for Lead Tungstate crystals of CMS EE quality and dimensions, superposed to values for EB type crystals from~\cite{r-CEFNIM}.}
\label{f-cumu}
\end{figure}
The major concern for HL-LHC running is that the energetic hadron fluences from p-p collisions are predicted to cause a specific, permanent, cumulative loss of
transmission in Lead Tungstate~\cite{r-LTNIM,r-pionNIM, r-fissionNIM}. Since a loss of transmission causes a reduction and a non-uniformity of light output that may worsen the energy resolution up to a level inadequate for HL-LHC running, the CMS ECAL group has decided to pursue this 
issue through in-beam studies of hadron-irradiated crystal matrices, the development of Monte Carlo (MC) simulations, to extrapolate the performance evolution through HL-LHC running, and through dedicated studies of crystal damage recovery through light injection and thermal treatment~\cite{r-SCINT11,r-SING}.
\begin{figure}[!t]
\centering
\includegraphics[width=3.5in]{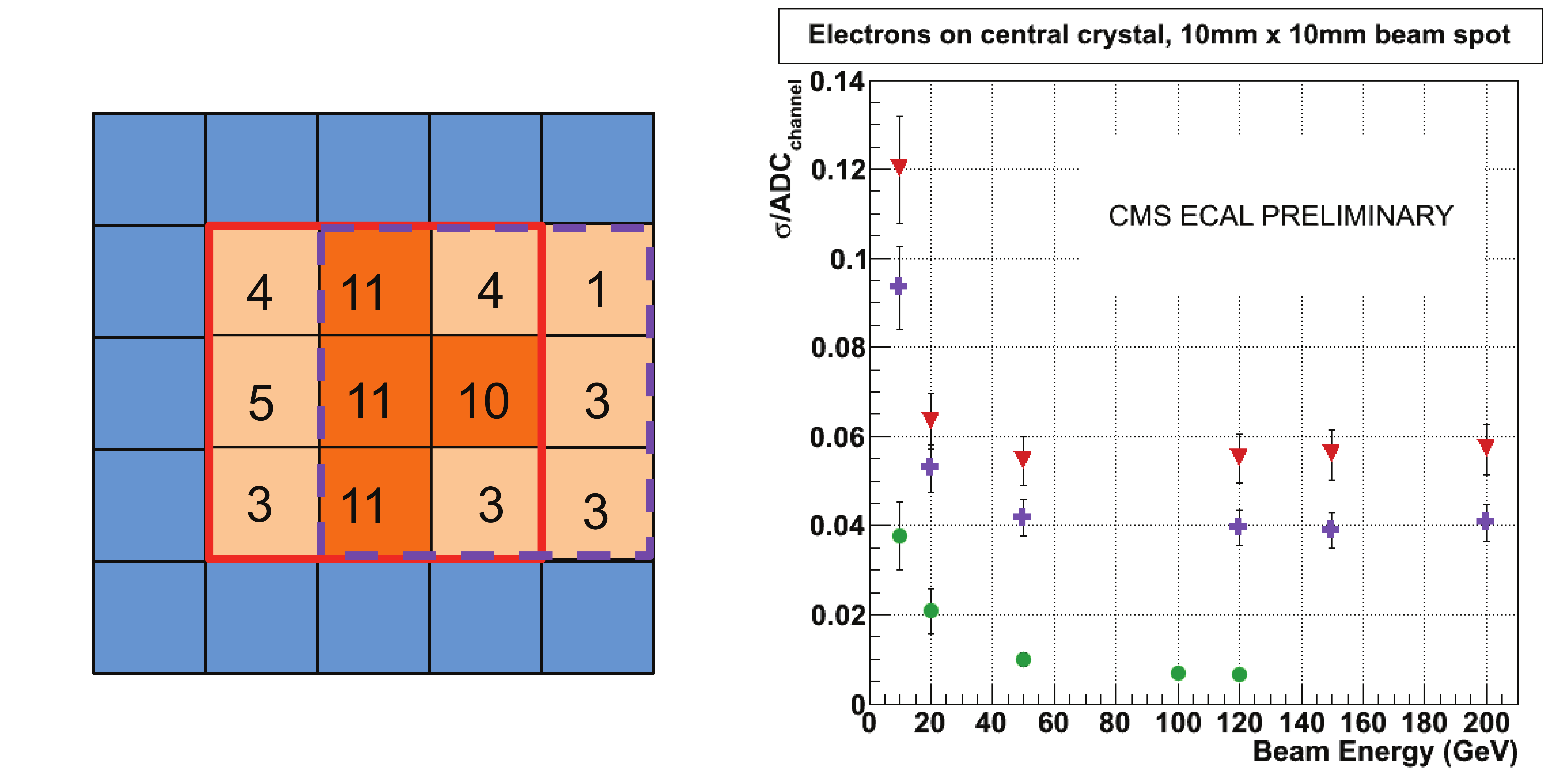}
	\caption{Left: Layout of the crystal matrix tested in electron beams at CERN. Values:  $\mu_{\mathrm{IND}}\; [\mathrm{m}^{-1}]$ for the proton-irradiated crystals. The red, continuous frame and the purple, dashed frame outline the $3\times3$ arrays for which the energy resolution was measured, with the beam hitting the central crystal. Right: Preliminary energy resolution as a function of beam energy for the two indicated arrays, where the color of the symbols matches the frame color on the left, compared to results for a non-irradiated matrix (round symbols). }
\label{f-Eres}
\end{figure}
\subsection{In-beam studies of hadron-irradiated crystal matrices}
 Lead Tungstate crystals of EE dimensions, and of a quality fulfilling the Technical Specifications for
CMS construction, were irradiated up to various integrated fluences with 24 GeV/c protons. The induced absorption values
observed are in agreement with results obtained for EB type crystals~\cite{r-CEFNIM} from two producers,
the Bogoroditzk Techno-Chemical Plant (BTCP) and the
Shanghai Institute of Ceramics (SIC), as shown in Fig.~\ref{f-cumu}.
These crystals were inserted into a $5\times5$ matrix of Lead Tungstate crystals, using a photomultiplier read-out with bi-alkali photocathodes, for studies in high-energy particle beams at the CERN SPS, as already described in~\cite{r-SING}. 
The energy resolution for electrons has been measured as a function of energy for these proton-irradiated matrices, and
compared to the results for a non-irradiated matrix. The crystals used in the matrices had been previously exposed to
proton fluences up to $\Phi_p = 6\times 10^{13}\;\mathrm{cm}^{-2}$, yielding induced absorption values as high as $11\;\mathrm{m}^{-1}$\footnote{smaller values for some crystals are due to a partial annealing of damage},
which correspond to $\lvert\eta\rvert=2.8$ after 500 $\mathrm{fb}^{-1}$ in the CMS ECAL~\cite{r-LTNIM,r-RADQ}. The layout of the two matrix configurations studied is visible in Fig.~\ref{f-Eres}.

The preliminary electron energy resolution as a function of beam energy for an array of $3\times 3$ non-irradiated crystals (round symbols), and for two arrays of $3\times 3$ proton-irradiated crystals in the matrix (same color for data points and for the corresponding array frame), for a central impact within 10 mm $\times$ 10 mm are shown in Fig.~\ref{f-Eres}.
At this time, the analysis method and results are still preliminary.
The data are uncorrected for irradiation fluence uncertainties, while the pedestal noise has been subtracted and the channel intercalibration for $3\times 3$ clusters has been performed at 50 GeV. The total energy has been reconstructed using a
$3\times 3$ cluster, by fitting the corresponding spectrum with the Crystal Ball function~\cite{r-CB}. The ordinate in the plot corresponds to the sigma of the crystal ball function over the peak position. The values obtained indicate that the energy resolution is clearly degraded with respect to a non-irradiated matrix, and can be used to benchmark a simulation aimed at extrapolating the performance evolution of the ECAL for different values of pseudorapidity and integrated luminosity. Furthermore, the difference in results between the two hadron-damaged $3\times 3$ arrays is clearly correlated with the difference in average induced absorption value.

\subsection{MC simulations for performance extrapolations}
To extrapolate the CMS ECAL performance to CMS running, a simulation has been developed, that includes ray-tracing in media
of given transmission, expressed through its absorption coefficient as a function of light wavelength.
The simulation has been developed using GEANT4~\cite{r-G4} to describe the detector geometry, the energy deposition mechanism and shower development, and sLITRANI~\cite{r-LIT} for ray-tracing. This section is intended to illustrate the successful benchmarking of the simulation framework using various sets of data. The work is still in progress, and after a full validation, the
framework will be used to evaluate the
evolution of ECAL performance that can be anticipated for HL-LHC running and to perform detector upgrade studies.

The first validation of the simulations with data is found in the correlation between Light Output loss and induced absorption
$\mu_{IND}$ after proton irradiation for two different sets of Light Output measurements.
\begin{figure}[!t]
\centering
\includegraphics[width=3.5in]{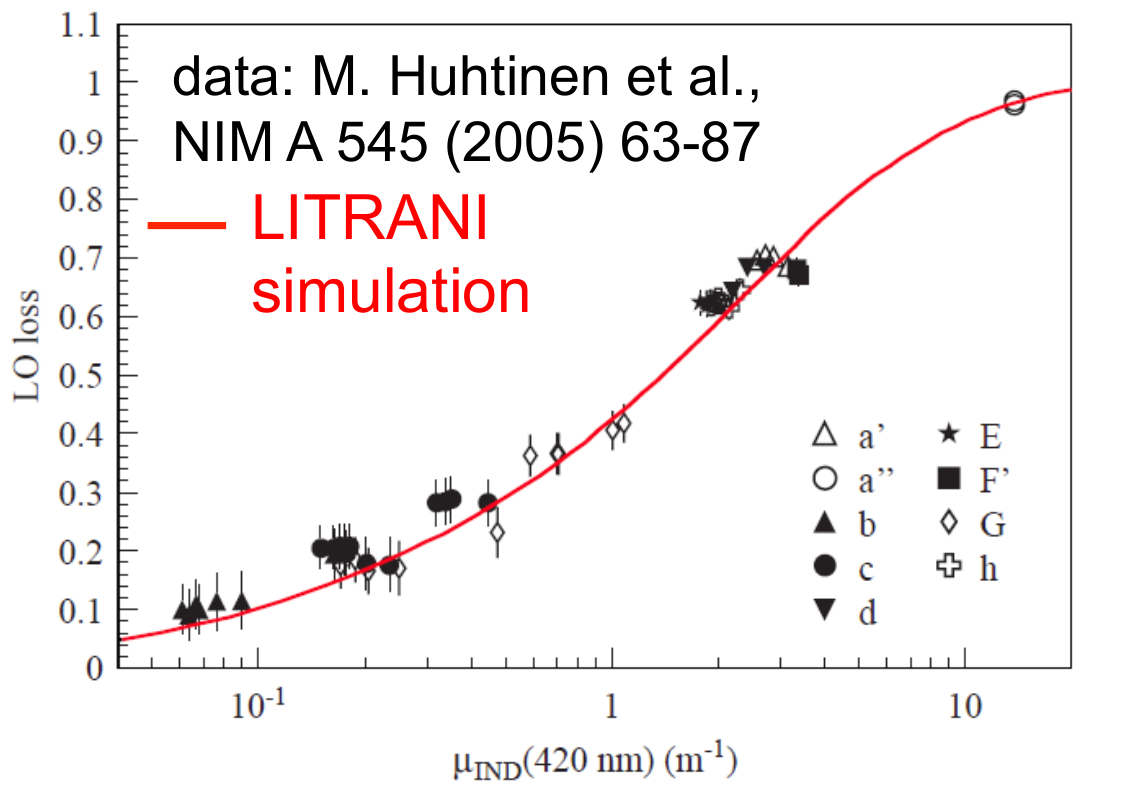}
\caption{Correlation between Light Output loss and  $\mu_{IND}$ (details in the text) for cosmic muons traversing proton-irradiated Lead Tungstate crystals laterally (symbols)~\cite{r-LYNIM}, compared to the interpolated simulation results (red line).}
\label{f-LOSSmu}
\end{figure}
In both sets, for the light detection, a photomultiplier has been used, with a bialkali photocathode covering the entire end face of the crystal. However, crystal type and the direction of the particles exciting scintillation were different in the two setups.

For the first set of data (Fig.~\ref{f-LOSSmu}), cosmic muons were selected using two trigger counters, as described in~\cite{r-LYNIM}.
A sLITRANI~\cite{r-LIT} simulation has been performed to obtain the light collection efficiency (EFF) versus distance Z from the crystal front face for a given value of $\mu_{IND}$. Scintillation photons have been simulated as emitted isotropically, with timing and wavelength properties as in~\cite{r-ECALTDR}.
Undamaged crystals are attributed a typical absorption length versus wavelength as measured for EE crystals, which is then applied uniformly inside a crystal.
The simulation of the average distribution of cosmic muons along the crystal, N(Z), has been performed using the dimensions of the experimental setup~\cite{r-LYNIM}. The distribution is not uniform, it has a maximum in the center of the crystal, and sharply decreases towards both ends of the crystal. The overall Light Output versus $\mu_{IND}$ is then obtained as
\begin{displaymath}
\int \mathrm{EFF}(\mathrm{Z}, \mu_{IND})\times \mathrm{N}(\mathrm{Z}) \mathrm{dZ},
\end{displaymath}
with an excellent agreement observed between measurements and simulation, for all values of $\mu_{IND}$.
\begin{figure}[!b]
\centering
\includegraphics[width=3.5in]{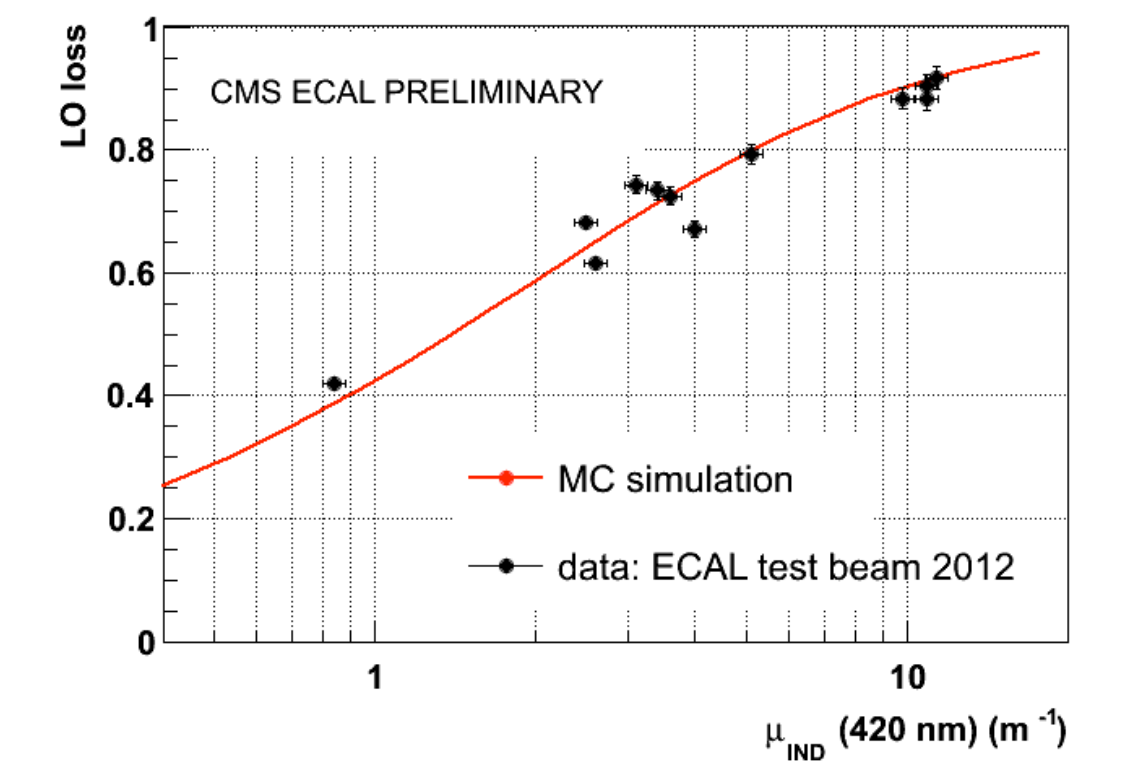}
	\caption{Correlation between Light Output loss and  $\mu_{IND}$ (details in the text) for 50 GeV/c test beam electrons hitting the proton-irradiated Lead Tungstate crystals front face (symbols), compared to the interpolated simulation results (red line).}
\label{f-LOSSe}
\end{figure}
\begin{figure}[!t]
\centering
\includegraphics[width=3.5in]{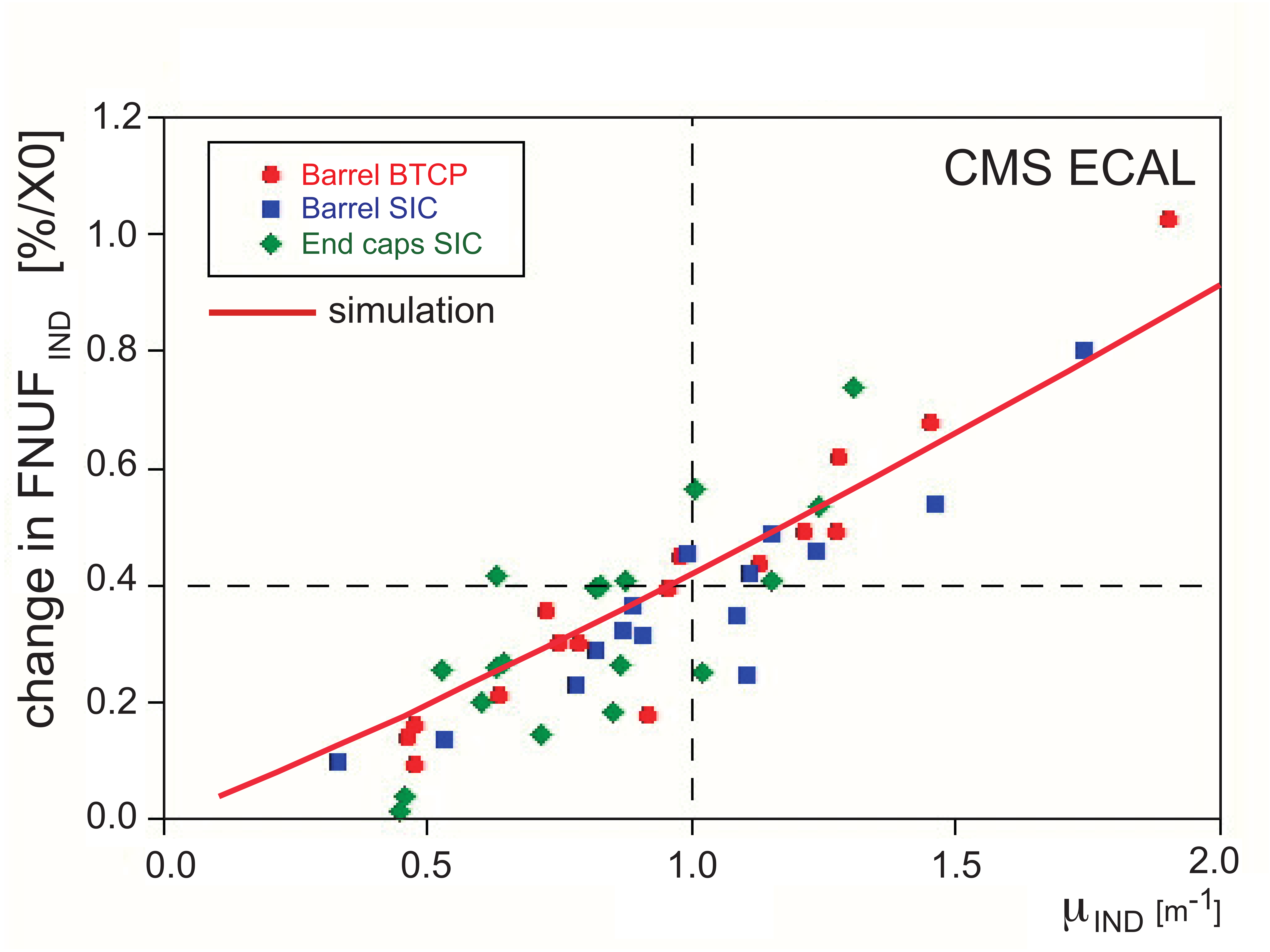}
	\caption{Correlation between FNUF and  $\mu_{IND}$ (details in the text) for measurements on three types of 
	Lead Tungstate crystals (symbols), compared to interpolated simulation results (red line).}
\label{f-FNUF}
\end{figure}

In the second set of data (Fig.~\ref{f-LOSSe}), the Light Output loss has been measured using high-energy electrons entering the front face of the crystals. The simulation with sLITRANI has been performed similarly as in the first case, and here as well, data and Monte Carlo agree.

A further validation is found in the correlation, after a $\gamma$  irradiation, between $\mu_{IND}$ and the change of uniformity in light output, quantified through the parameter FNUF, the slope of a linear fit applied to the Light Yield measurements as a function of the longitudinal position between 13 $X_0$ and 4 $X_0$. It is expressed in relative change per radiation length~\cite{r-FNUF}.
This is a parameter that crucially contributes to the energy resolution~\cite{r-BEAU}.
The measurements are shown in Fig.~\ref{f-FNUF}, superposed to the simulation results. 
For the simulation, the light collection efficiency (EFF) versus distance Z from the front face of EE crystals is calculated with
sLITRANI ~\cite{r-LIT}. 
The procedure of FNUF determination is identical to the one used for crystal Quality Control during ECAL construction, and it is applied to EFF vs Z for a given induced absorption   $\mu_{IND}$.
The difference $FNUF(\mu_{IND}=0) - FNUF (\mu_{IND})$ versus  $\mu_{IND}$ is plotted as a red line. 
A good agreement is observed between measured and simulated values.

\begin{figure}[!b]
\centering
\includegraphics[width=2.8in]{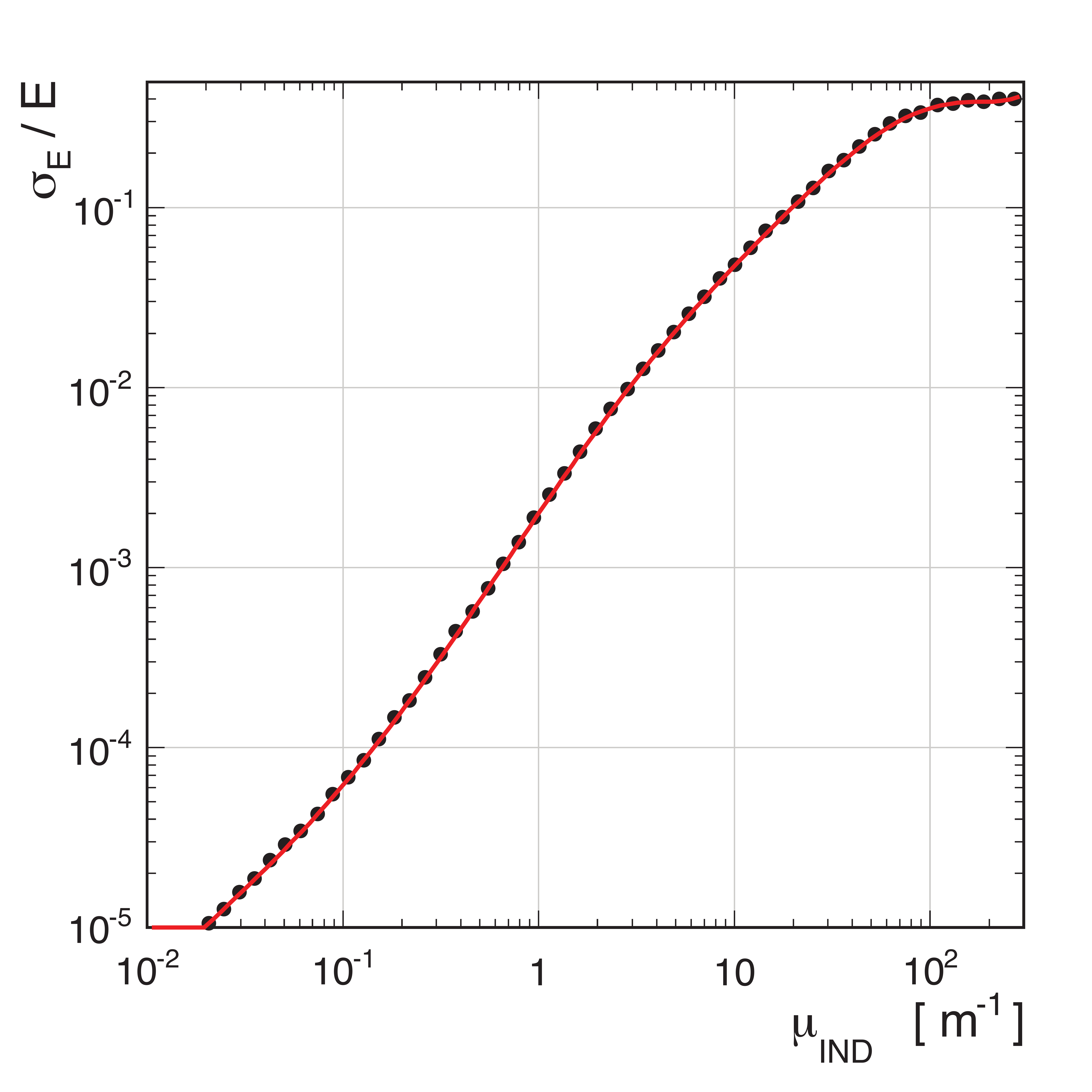}
	\caption{Simulation results for the contribution to the EE energy resolution as a function of $\mu_{IND}$ (details in the text)
	from longitudinal non-uniformity alone.}
\label{f-MCRES}
\end{figure}
The contribution to energy resolution due to the longitudinal non-uniformity in light collection caused by a loss of crystal  transmission is the third validation performed. The simulation
results are shown in Fig.~\ref{f-MCRES}, where it can be observed that, for a typical value $\mu_{IND}\simeq 10\;{\mathrm m}^{-1}$ as present in the test beam matrix of Fig.~\ref{f-Eres}, the predicted contribution, $\frac{\sigma_E}{E}\simeq 0.05$, is consistent with the change of energy resolution between an unirradiated matrix and one containing hadron-irradiated crystals, thus offering a further validation of the simulation framework.

\section{Outlook}
Measurements have been performed, to asses the changes in ECAL energy resolution expected from hadron-damage to crystals, and a Monte Carlo simulation has been validated, to extrapolate the performance to the running conditions
envisaged for High-Luminosity running.

The need for an ECAL EE upgrade for HL-LHC running will be assessed through physics performance studies, and depending 
on their outcome, the design parameters will be optimized accordingly. It is clear that the technology chosen would need to
withstand the hostile environment, while remaining affordable and practical.
Various options for calorimetry at the HL-LHC are being studied in detector simulations and R\&D work, as described through some examples in the following section.

\section{R\&D work on calorimetry at the HL-LHC}
\label{RnD}
Several options are being studied, for the case that an upgrade of the ECAL end caps became necessary for HL-LHC running, as mentioned already in~\cite{r-CMSUTDR}. 
Ongoing R\&D work in collaborating institutes concern scintillating crystals such as LYSO~\cite{r-MAO,r-FNLYSO}, YSO, LuAG~\cite{r-LuAG}, Cerium Fluoride~\cite{r-CEFNIM}, while the design options considered include homogeneous calorimetry, as well as a sampling calorimeter.
It is worth mentioning here, that some scintillators have already been identified, that are extremely radiation hard, while mass production and cost issues are further elements to be taken into account.

A further object of investigation are scintillating ceramics, which are a new, promising and ongoing development in the field of inorganic scintillators. As sintered materials, their advantages compared to crystals lie in the lower temperatures needed for production, the faster production cycle, the possibility to be formed to shape and to be machined easily, due to the absence of cleavage planes.
Where scintillation is induced by an activator, its uniformity is intrinsic to the production method. Drawbacks can be found in the fact that transparent materials can be obtained successfully through a sintering process only for a cubic crystalline structure, and in the fact that this field of material research has just started giving first results.
We consider such a material, at the present stage of development, potentially suitable for a sampling calorimeter option. This R\&D is being pursued in some of the institutes participating in the CMS ECAL and in laboratories that develop such materials, see e.g.~\cite{r-CER}.

\section{Conclusions}
The present evolution of CMS ECAL response is found to be in agreement with the technical specifications applied during
construction. 
The contributions from hadron-specific damage to signal losses is expected to become significant starting at the highest $\eta$ values in the endcaps, towards the end of LHC running. Beam test results on irradiated crystals have been used to quantify the effects expected from cumulative hadron damage. Work is in progress to disentangle the various contributions to the already observed signal losses, and some qualitative results are already available. Simulations have been developed that incorporate signal losses and their effects on energy resolution, and their benchmarking has been successful.
While physics needs are being studied, R\&D for HL-LHC running is being performed in parallel, in collaborating ECAL institutes.

\end{document}